\documentstyle[12pt]{article}
\renewcommand{\theequation}{\arabic{section}.\arabic{equation}}
\begin{document}

\topmargin=-2.0cm

\evensidemargin=0cm
\oddsidemargin=0cm
\date{}
\baselineskip=0.7cm

\renewcommand{\appendix}{\renewcommand{\thesection}
{Appendix}\renewcommand{\theequation}{\Alph{section}.\arabic{equation}}
\setcounter{equation}{0}\setcounter{section}{0}}

\vspace*{-2cm}
\begin{flushright}UT-Komaba 96-6\\
hep-th/9604032
\end{flushright}
\vspace{1.cm}

\begin{center}
{\LARGE {\bf Boson Expansion Methods\\ }}
\vspace{5mm}
{\LARGE {\bf in (1+1)-dimensional Light-Front QCD}}\\
\vspace{5mm}

\vspace{1.5cm}
{\Large Kazunori Itakura}\footnote{Electronic address: 
itakura@hep1.c.u-tokyo.ac.jp}\\

\vspace{0.7cm}

{\it Institute of Physics, University of Tokyo,}\\
{\it Komaba, Meguro-ku, Tokyo 153, Japan}

\vspace{0.7cm}

{\bf Abstract}
\end{center}
We derive a bosonic Hamiltonian from 
two dimensional QCD on the light-front.
To obtain the bosonic theory we find that it is useful 
to apply  the boson expansion method 
which is the standard technique in quantum many-body physics.
We introduce bilocal boson operators 
to represent the gauge-invariant quark bilinears and 
then local boson operators as the collective 
states of the bilocal bosons.
If we adopt the Holstein-Primakoff type  
among various representations, 
we obtain a theory of  infinitely many interacting bosons,
whose masses are  the eigenvalues of the 't Hooft equation.
In the large $N$ limit,  since the interaction disappears and 
the bosons are identified with mesons,  
we obtain a free Hamiltonian
with infinite kinds of mesons.\\

PACS number(s); 11.10.Kk, 11.10.St, 11.15.Tk, 11.15.Pg

\vspace{1.5cm}
\begin{center}
{\it To be published in Physical Review D}
\end{center}

\newpage

%%%%%%%%%%%%%%%%%%%%%%%%%%%%%%%%
%%%%%%%               INTRODUCTION            %%%%%%%%
%%%%%%%%%%%%%%%%%%%%%%%%%%%%%%%%
\section{Introduction}

The low energy dynamics of  strong interactions is usually 
investigated in the framework of  effective theories introduced 
phenomenologically \cite{Donoghue}.
Now that the strong interaction is believed to be described by
quantum chromodynamics (QCD), 
these effective theories, if they are  correct,  
should be derived from QCD in some way or another.  
For example, although the chiral perturbation theory 
or the Skyrme model describes the low energy physics
involving the Nambu-Goldstone (NG) bosons,
we cannot determine definitely the forms of  interactions 
or the magnitude of the parameters.
Then a natural question arises of how to obtain directly
such a hadronic effective theory from QCD, 
which may be answered after  solving  the problem 
of confinement and dynamical chiral symmetry breaking.
For the realistic case, QCD${}_{3+1}$, 
this problem is still far beyond our knowledge.
Two-dimensional QCD, however,  may serve as a tractable model 
at least to know some consequences after confinement.

Two dimensional QCD occupies  a very unique position 
as a toy model for the 4 dimensional QCD.
Originally it was introduced by 't Hooft \cite{tHooft}
to demonstrate that we can sum up 
the planar diagrams and can get  nonperturbative results 
in the limit of a large number of colors (the large $N$ limit).
The Bethe-Salpeter equation for the quark-antiquark system 
becomes simple and leads to 
what is now called  the 't Hooft equation  which  determines  
the mesonic spectra in the large $N$ limit.
Second, two dimensional gauge theories have a
linear potential and, thus, are 
 confining in the lowest perturbation theory. 
Also, nonperturbatively,  't Hooft showed that 
the mesonic spectra are discrete,
which suggests the confinement of  quarks 
in the large $N$ limit.
Third,  Kikkawa \cite{Kikkawa}
 presented the idea that the Hamiltonian 
can be easily rewritten only in terms of a 
 gauge-invariant bilocal operator 
due to the absence of  propagating 
degrees of freedom in two dimensional gauge theories.
This  operator is a path-ordered  product  
with its ends attached by  a quark and an anti-quark, 
and thus physically close to a mesonic state.
Several works using  gauge-invariant bilocal operators
 have been performed in two dimensional QCD 
\cite{Naka-Oda,Rajeev,Wadia1,Wadia2,Cavicchi,Barbon}. From 
these facts, two dimensional QCD can be understood 
as an appropriate model to investigate  how hadronic theory 
can be constructed after confinement and to discuss the 
higher order effects of the $1/N$ expansion,
which is necessary for the realistic case $N=3$.

Since both of the works above  greatly benefited from 
the light-front (LF) frame,
we also work on the light-front to go beyond them.
Recently the light-front Hamiltonian field theory 
has acquired renewed interest 
with the hope of solving  nonperturbatively the 
relativistic bound-state problems
\cite{Dirac,Wilson,Perry,Zhang,Burkardt}.
Its main strategy is the "trivial" vacuum 
and the approximation to a restricted Fock space 
with a few number of particles.
Bound states are analyzed
within a subspace with a few numbers of particles. 
In this sense, this approach is similar to
 the constituent quark picture, where  
the dynamical variables are a few constituent quarks.     
On the LF, a careful treatment of the zero mode  
is important for an understanding of  
spontaneous symmetry breaking 
(for example, see \cite{MasYama,Pinsky}) and 
 vacuum structures such as 
the $\theta$ vacuum, 
but we do not consider the zero modes 
in this paper and we will 
work on the infinite  light-front space.
This is reasonable because in two dimensions 
there is no spontaneous symmetry breaking 
for {\it finite} $N$
 theories \cite{Coleman} and also 
there is no topological configuration
in $SU(N)$ or $U(N)$ QCD${}_{1+1}$ 
with fundamental fermions. 
According to the calculations of the various 
two-dimensional models
\cite{Bergknoff,MoPerry,HSTM,HOT,Hornbostel,Sugihara,Kallo}, 
we know that the Tamm-Dancoff approximation 
on the light-front 
(Light-Front Tamm-Dancoff approach, LFTD)
\cite{Tamm,Dancoff,LFTD}
 is  very good for lower excitations.
Especially the lowest meson state is 
found to be  a two-body state  
(a quark-antiquark state) almost perfectly.
Although the success  of LFTD will be partially 
due to the triviality of the vacuum,
why LFTD is so good is not  thoroughly 
understood yet and it is not certain  that 
we can also describe  the NG bosons
in this approximation. 
Indeed, in the constituent quark model, 
pions cannot be described by few constituents.
Since there are no NG bosons in two dimensions, 
the success of LFTD in two dimensions 
does not tell anything about the NG bosons.
Furthermore, a more fundamental question of 
whether we can describe  the NG bosons on the LF
 is not so clearly answered to date \cite{Tsuji}.
Therefore approaches independent 
of the Tamm-Dancoff approximation 
will be needed to understand why LFTD is good 
and to go to higher dimensions where the NG mode may exist.

Motivated by these we introduce  a method to obtain a 
bosonic Hamiltonian from fermionic systems.
This  is  a well-known technique in quantum many-body physics
and  is called  the boson expansion method 
or the boson mapping \cite{Marshalek}.
Using this method we can introduce bosonic variables to represent 
bifermion operators. On the other hand, 
the work of Kikkawa can be understood as the operator theory
of the quark-antiquark bound states.
The variable in the formalism is the bilocal operators.
We extract bosonic variables from the bilocal operators and 
construct a bosonic theory.
Since the main purpose of this paper is to indicate that 
the use of the boson expansion method is 
very natural to obtain a bosonic theory on the LF, 
 its conceptual aspects are emphasized throughout the paper.

This paper is organized as follows.
In the next section, 
the basics of QCD$_{1+1}$ on the LF is presented
and the gauge-invariant approach by Kikkawa is given.
It is shown that an equation of motion for the two-body state
 becomes the 't Hooft equation in the large $N$ limit.
In Sec. III, we consider the general structures
 of the two-body states
and why we need boson expansion methods is explained.
In Sec.  IV, the boson expansion methods 
are shortly reviewed and applied to 2D QCD. 
Here we introduce bilocal boson operators 
to represent fermion bilinear operators.
Local boson operators are given in Sec. V 
as collective bosons of the bilocal operators.
Finally  Sec. VI is devoted to a summary and discussion. 
The Appendix shows  some relations for  operators between the 
gauge-invariant approach and the usual one.

%%%%%%%%%%%%%%%%%%%%%%%%%%%%%%%%
%%%%%%%         Bilocal Formulation               %%%%%%%%
%%%%%%%%%%%%%%%%%%%%%%%%%%%%%%%%
\section{Light-front QCD$_{1+1}$ and its bilocal formulation}
\setcounter{equation}{0}

In this section, we fix the notation for two dimensional QCD 
on the light front and introduce its bilocal formulation of Kikkawa 
\cite{Kikkawa}.
This formalism gives a Hamiltonian  
only in terms of gauge-invariant bilocal operators, 
which are essentially quark bilinears. In the large $N$ limit,
the Heisenberg equation of motion for the bilinear operator 
becomes  the 't~Hooft equation 
\cite{tHooft}.

\subsection{QCD$_{1+1}$ in light-front coordinates}

The model we consider is a two-dimensional gauge theory
 with a massive quark 
in the fundamental representation of $SU(N)$ or $U(N)$. 
For simplicity we treat a one-flavor quark, but the generalization 
to many flavors is straightforward.
The Lagrangian density is 
\begin{equation}
{\cal L}=-\frac{1}{2} {\rm tr} 
(F_{\mu\nu}F^{\mu\nu})+\bar \Psi(i\gamma^{\mu}D_\mu-m)\Psi,
\end{equation}
where $F_{\mu\nu}=\partial_\mu 
A_\nu -\partial_\nu A_\mu -ig [A_\mu, A_\nu]$
is the  field strength and $D_\mu=\partial_\mu-igA_\mu
=\partial_\mu-igA^a_\mu T^a$
 is a covariant derivative. The $N\times N$ matrices $T^a$, 
generators of the group, 
are normalized as ${\rm tr} (T^aT^b)=
\frac{1}{2} \delta^{ab}$ and satisfy 
the Lie algebra $ [T^a, T^b]=if^{abc}T^c$. 
The fundamental quark $\Psi$ is a 
two-component spinor and we define 
\begin{equation}
\Psi_i=2^{-\frac14}\left(  \begin{array}{c}
				    \psi_i  \\ \chi_i   
				\end{array}	\right),\ \ \
\end{equation}
where $i=1,\ldots,N$ is a color index.
We use the following $\gamma$-matrices;
$
\gamma_0=\sigma_1,\ 
\gamma_1=i\sigma_2.
$

We work in the light-front coordinates,
$x^{\pm}=(x^0\pm x^1)/\sqrt{2},$ 
and thus regard $x^+$ as 'time'
 and so $\partial_+=\frac{\partial}{\partial x^+} (=\partial^-) $ 
as 'time' derivative. 
To avoid  confusion, we shall use only $\partial_+$ 
and the spatial derivative 
$\partial_-$ instead of $\partial^-$ or $\partial^+$. 
If we choose the light-front 'temporal' gauge, $A^+=0$,
the Lagrangian becomes 
\begin{equation}
{\cal L}={\rm tr} (\partial_- A_+)^2 + i\psi^{\dagger } D_+ \psi 
           + i \chi^{\dagger }\partial_-\chi
           -\frac{m}{\sqrt{2}}(\psi^{\dagger }\chi 
           + \chi^{\dagger }\psi).\label{lag}
\end{equation}
It is evident that $A_+$ and $\chi$ are not dynamical,
because there is no kinetic term for them in eq.(\ref{lag}).
The constraints,
\begin{eqnarray}
\partial_-^2 A_+^a-g\psi^\dagger T^a \psi &=&0,\\
i\partial_-\chi-\frac{m}{\sqrt{2}}\psi&=&0,
\end{eqnarray}
are solved easily and the solutions are 
\begin{eqnarray}
A_+^a(x)&=&\frac{g}{2}\int_{-\infty}^{\infty}
 dy\ |x-y| J^a(y) + C^ax+ D^a,\\
\chi(x)&=&\frac{m}{2\sqrt{2}i}
\int_{-\infty}^{\infty} dy\ \epsilon(x-y) \psi(y),
\label{fermisol}
\end{eqnarray}
where $J^a(x)=\psi^{\dagger}(x)T^a\psi(x)$ and $\epsilon(x)$
is a sign function $\epsilon(x)=x/|x|$.\footnote{ From 
now on we write the light-front spatial variables 
without their suffices, i.e. $x$ and $p$ instead of $x^-$ or $p_-$.
Moreover, since we work in the Hamiltonian formalism,
we omit the light-front 'time' coordinate.}
As was pointed out by Bars and Green \cite{BarsGreen}, 
we can set $C^a=D^a=0$ in the physical space. 
And also, since we do not consider the fermionic zero mode,
we did not write the boundary terms in eq. (\ref{fermisol}).
Inserting these solutions into the Hamiltonian and using 
the same trick for the 
product of the sign functions as that of Kikkawa \cite{Kikkawa}, 
we have 
\begin{eqnarray}
H&=&-\frac{g^2}{4}\int dxdx'\ |x-x'|J^a(x)J^a(x')\nonumber  \\
&&+\frac{m^2}{4i}\int dx dx'\ 
\epsilon(x-x')\psi^\dagger(x)\psi(x') .\label{orgH}
\end{eqnarray}
The remnant of the gauge field can be found 
as the linear potential in the first term.
Finally,  a quantization condition is imposed 
on the dynamical variable $\psi$;
\begin{eqnarray}
\{\psi_i(x), \psi^{\dagger}_j(y)\}_{x^+=y^+}&=&
\delta_{ij}\delta(x-y),\nonumber  \\
\{\psi_i(x), \psi_j(y)\}_{x^+=y^+}
&=&\{\psi^{\dagger}_i(x), \psi^{\dagger}_j(y)\}_{x^+=y^+}=0,\label{anticom}
\end{eqnarray}
The Fock vacuum is 
\begin{equation}
a^i_{k_-}|0>=d^i_{k_-}|0>=0 \hspace{1cm} 
(\forall i=1,\ldots,N \ \ , \ k_->0),\label{vac}
\end{equation}
where annihilation operators are defined as
\begin{equation}
\psi_i(x)=\int_0^\infty \frac{dk_-}{\sqrt{2\pi}}
\left\{ a^{i}_{k_-} e^{-ikx}+d_{k_-}^{i\dagger} e^{ikx} \right\}.
\label{usualrep}
\end{equation}
On the LF, as long as we do not consider the zero mode,
the Fock vacuum $|0>$ is  also the  vacuum of the 
interacting Hamiltonian (\ref{orgH}).

\subsection{Bilocal formulation--Gauge invariant variables}

Following Kikkawa \cite{Kikkawa}, 
let us introduce  a color-singlet 
bilocal field and its Fourier transform;
\begin{equation}
M(x,y)=\frac{1}{2}\sum_{i=1}^N [\psi^\dagger_i(x), \psi_i(y)], 
\label{bilocal}
\end{equation}
\begin{equation}
M(p_1,p_2)=\int_{-\infty}^{\infty}\frac{dx_1}{\sqrt{2\pi}}
\frac{dx_2}{\sqrt{2\pi}}\ \ M(x_1,x_2)e^{ip_1 x_1+ip_2 x_2}.
\label{bilocal_mom}
\end{equation}
Although this is defined as the commutator of fermion fields, 
it is not essential because we treat only normal ordered 
operators in the following discussions.
If we choose the $A^-=0$ gauge as in Ref. \cite{Kikkawa},
$M$ is given by
\begin{equation}
{\cal M}(x,y)=\frac{1}{2}\left[ \psi^{\dagger}(x)
{\rm P}e^{ig\int^x_y A_-(z)dz^-}\ ,\ \ 
\psi(y)\right],\label{bilocal2}
\end{equation}
where P denotes the path-ordered product.
Since we are working at a fixed time in two dimensions, 
there is no path dynamics for the Wilson line. 
This operator is gauge equivalent to Eq.(\ref{bilocal}).
Indeed if we perform the gauge transformation,
$U(x)={\rm P}\exp \{ig\int_x^{-\infty}A_-dz^-\}$, 
 Eq.(\ref{bilocal2}) is transformed into Eq.(\ref{bilocal}).

Quantization of $M(p_1,p_2)=M^\dagger(p_2,p_1)$ 
 is given by the  commutation relation;
\begin{eqnarray}
&&\hspace{-1.5cm}[:M(p_1,p_2):,:M(q_1,q_2):]\nonumber  \\
&=&:M(p_1,q_2):\delta(p_2+q_1)-:M(q_1,p_2):\delta(p_1+q_2)
\label{Algebra}\\
&+&N\delta(p_1+q_2)\delta(q_1+p_2)
		\{ \theta(p_1)\theta(p_2)\theta(-q_1)\theta(-q_2)
-\theta(-p_1)\theta(-p_2)\theta(q_1)\theta(q_2)\}.\nonumber
\end{eqnarray}
This  is derived from the anti-commutation relations (\ref{anticom}).
The normal order of $M$ is defined 
through the quark operators (\ref{usualrep}) 
in terms of the Fock vacuum.
The last c-number term that emerged from normal-ordering  
is very important for later discussions.
Decomposing $M$ into four parts by the sign of the momentum as
\begin{equation}
M_{\tau_1\tau_2}(p_1,p_2)=
\theta(\tau_1p_1)\theta(\tau_2p_2)M(p_1,p_2),
\end{equation}
we can see that only $:M_{--}:$ does not annihilate the vacuum;
\begin{equation}
:M_{++}(p_1,p_2):|0>=:M_{+-}(p_1,p_2):|0>=:M_{-+}(p_1,p_2):|0>=0.
			\label{vac2}
\end{equation}
The state $:M_{--}:|0>$ corresponds to
 the quark-antiquark two-body state.\footnote{We call this state 
as  a "two-body state" instead of a "mesonic state"
as in Ref. \cite{Kikkawa}, 
because this is not an eigenstate of the Hamiltonian
as we shall see. }
Although the full Fock space is constructed 
by further introducing gauge-invariant baryon operators \cite{Kikkawa},
 we  treat, in this paper,  only the bosonic sector ( i.e. a 
subspace with a zero fermion number );
\begin{equation}
{\cal F}_B=\left\{ \prod_i \left(:M_{--}(p_i,q_i):\right) |0>\right\}.
\end{equation}
The Hamiltonian (\ref{orgH}) is rewritten 
by using the normal ordered bilocal fields as
\begin{eqnarray}
H&=&-\frac{g^2}{8\pi}\int_{-\infty}^{\infty}dp_1dp_2dq_1dq_2\ 
\delta(p_1+p_2+q_1+q_2)\times \nonumber  \\
&&\hspace{1cm}\left\{\frac{{\cal P}}{(p_1+q_2)^2}+
\frac{1}{N}\frac{{\cal P}}{(p_1+p_2)^2}
\right \}:M(p_1,p_2)::M(q_1,q_2):\nonumber  \\
&&-\frac{1}{2}\left(m^2-\frac{g^2N}{2\pi}\right)
\int_{-\infty}^{\infty}dq\ \frac{{\cal P}}{q}:M(q,-q):,\label{Ham}
\end{eqnarray}
where ${\cal P}$ denotes the principal value prescription 
of the integral to avoid infrared divergences \cite{tHooft}.
The  term with $1/N$ in the first term 
is absent for the $U(N)$ theories.
This Hamiltonian is not completely normal ordered, e.g. 
 $:M_{+-}::M_{--}:$ is not in the normal order.  
This is because we want to keep in touch with 
the unit $:M:$ which is close to the mesonic object.
Although  the Hamiltonian (\ref{Ham}) is quadratic 
in terms of $:M:$,  the Heisenberg equation of motion 
for the two-body state $:M_{--}:|0>$ 
has interaction terms due to the nontrivial 
algebra (\ref{Algebra}).
For $U(N)$ theory 
(hereafter we treat only $U(N)$ case for simplicity) 
we obtain  
\begin{eqnarray}
&&\hspace{-2cm}i\partial_+:M_{--}(q_1,q_2):|0>
=[:M_{--}(q_1,q_2):,\ \ :H:]|0>\nonumber  \\
&=&\frac{g^2N }{4\pi}\int dk \frac{{\cal P}}{k^2}
:M_{--}(q_1-k,q_2+k):|0>\nonumber  \\
&+&\frac{1}{2} \left(m^2-\frac{g^2N}{2\pi}\right) 
\left(  \frac{1}{q_1}+
\frac{1}{q_2}  \right):M_{--}(q_1,q_2):|0> \nonumber  \\
&-& \frac{g^2}{4\pi}\int dkdp \ \frac{{\cal P}}{k^2}\big(
:M_{--}(q_1,k-p)::M_{--}(p,q_2-k):\nonumber  \\
&&\hspace{3cm}-:M_{--}(p,q_2)::M_{--}(q_1-k,k-p):     
 \big)|0>. \label{EOM}
\end{eqnarray}
Since it is reasonable to treat 
$:M_{\pm \pm}:$ and $:M_{\pm\mp}:$ 
to be $O(N^{1/2})$ and $O(N^0)$ respectively, 
we find that the first two terms give the leading 
contribution ($O(N^0)$) in the large $N$ limit 
($N\rightarrow \infty $ with $g^2N$ fixed), and that 
the last term is in the next order.
The last term represents the 
dissociation of a single two-body state into two two-body states.
The leading contribution  becomes a homogeneous equation 
and leads to the 't Hooft equation \cite{tHooft} which
determines the mesonic spectra in the large $N$ limit,
\begin{equation}
\mu^2 \phi(z) = \left( m^2-\frac{g^2N}{2\pi}\right)
\left( \frac1z + \frac{1}{1-z}\right) \phi(z)-
\frac{Ng^2}{2\pi}\int_0^1 dy \frac{{\cal  P}}{(y-z)^2}
\phi(y),\label{tHoofteq}
\end{equation}
where $\mu^2=2r_-r_+$ is the invariant  mass, 
$r_-=q_1+q_2 $ is the total momentum, 
$ z=q_1/r_-$ is the momentum fraction
of a quark,  and 
$\phi(z)=<0|(:M_{--}(q_1,q_2):)^{\dagger}|\phi>$ 
is the wave function of the relative motion.
Therefore the two-body state can be identified 
with a meson in the large $N$ limit.
The meson does not dissociate into 
two mesons in leading order.

%%%%%%%%%%%%%%%%%%%%%%%%%%%%%%%%%%
%%%%%%%%         TWO-BODY STATES           %%%%%%%%%
%%%%%%%%%%%%%%%%%%%%%%%%%%%%%%%%%%
\section{General consideration of two-body states}
\setcounter{equation}{0}
In the previous section, we  learned from Kikkawa's formalism 
that even though we start from the Hamiltonian (\ref{Ham}) quadratic
in terms of  the bilocal operators, the resulting  Heisenberg
equation of motion (\ref{EOM}) shows that there exists an interaction.
In the large $N$ limit, however,  the interaction vanishes and 
thus the two-body state becomes free and  is identified with the meson.
The origin of the difference between  finite  and  infinite $N$ theories
is that while the algebra (\ref{Algebra}) is nontrivial for finite $N$, 
it goes to a bosonic commutator in the large $N$ limit 
($:M_{++}:$ is the hermitian conjugate to $:M_{--}:$, see the Appendix);
\begin{equation}
\lim_{N\rightarrow\infty}[:M_{++}(p_1,p_2):,:M_{--}(q_1,q_2):]
=N\delta(p_1+q_2)\delta(p_2+q_1),\label{largeNB}
\end{equation}
thus the quadratic Hamiltonian gives 
a homogeneous equation of motion.
Therefore we insist that {\it
the interaction between two-body states 
has been hidden in the nontrivial algebra 
of the bilocal operators.}
What  we want to do is to construct 
an interacting meson theory for {\it finite} $N$.
It would be more convenient 
if we can see the interaction manifest in the Hamiltonian.
This will be achieved if we can express 
the two-body operators in terms of, 
say, some bosonic operators.
Using bosonic operators is justified as follows.

For finite $N$ theories,
 the real mesonic state  may contain,
 in general,  many-body components 
in addition to the two-body state;
\begin{equation}
|{\rm meson}>=|q\bar{q}>+|q\bar{q}q\bar{q}>
+|q\bar{q}g>+\cdots,
\end{equation}
where $|q\bar{q}>$ is a quark-antiquark two-body state, 
 $|q\bar{q}q\bar{q}>$ is a four-body state, and so on.
As was mentioned in the Introduction, 
in the context of the light-front 
Tamm-Dancoff approach \cite{Tamm,Dancoff,LFTD},  
however, we know that in various models 
\cite{Bergknoff}-\cite{Kallo} 
the lowest meson consists of  only a two-body component.
Thus to regard the two-body state as  a  meson 
would not be so bad at least 
for the lowest mesonic excitation 
even for finite $N$ theory.
A general (color singlet) two-body state with its total momentum $P$
 may be given as the linear combination of 
$:M_{--}(p,q):|0>$ with some normalization factor 
${\cal N} $,
\begin{eqnarray}
|q\bar{q}>&=&{\cal N} \sum_i \int_0^P dk_- 
\phi(k_-)a^{i\dagger}_{k_-}d^{i\dagger}_{P-k_-}|0>\nonumber  \\
&=&{\cal N}\int_0^P dk_-\phi(k_-):M_{--}(-k_-,-P+k_-):|0>,
\end{eqnarray}
where  $\phi(k)$ is 
the momentum space wave function of the relative motion 
between a quark and an antiquark (see the Appendix 
for the relation between $M$ and 
the quark or antiquark operators). 
Even if this state corresponds to the lowest meson, 
it is not a bosonic state because of 
the nontrivial algebra  (\ref{Algebra}).
We cannot construct bosonic operators 
only from quark-antiquark operators. 
For massless $U(1)$ gauge theory,
 a regularized current operator satisfies  bosonic commutators 
and it gives the usual bosonization; 
we obtain a Hamiltonian of a free massive boson. 
However, it contains $a^{\dagger}a$ and $d^{\dagger}d$ terms
and is not  a pure quark-antiquark operator.

The physical meaning of the non-bosonic properties
of the two-body state is given as follows.
Since the two-body state consists of fermions,  
the Pauli principle works and 
forbids the two-body states 
to be in the same state, for example 
the same momentum state. 
This effect works between two-body states.
Indeed a single two-body state can be considered 
as a bosonic state, which is observed as
\begin{equation}
<0|[:M_{++}(p_1,p_2):,:M_{--}(q_1,q_2):]|0>
=N\delta(p_1+q_2)\delta(p_2+q_1),
\label{singleB}
\end{equation}
for arbitrary $N$.
Hence the interaction between two-body states 
can be understood as  the effect of the Pauli principle 
and thus this is a universal property of the fermion-pair systems.
Note that the equation (\ref{largeNB}) can 
also be understood in the context  of the Pauli principle; 
when  $N\rightarrow \infty$, the number of allowed states 
for a fixed momentum goes to infinity and thus the Pauli principle 
loses its efficacy.

Since the two-body state is a good approximation for the lowest meson,
we want to regard the state as a bosonic state.
Indeed it behaves as a boson  in the large $N$ limit
 (\ref{largeNB}) and  even for finite $N$, 
as long as  there is a single two-body state (\ref{singleB}).
Therefore it is natural to try to find a representation 
where  $:M:$ is expressed by 
some bosonic operator which is equivalent  to
 the two-body state in these cases. 
If we are able to do this,  we obtain a bosonic theory
with interactions manifest   in the Hamiltonian.
As for the problem of how to represent operators 
whose algebra is nontrivial, 
in terms of bosonic operators, 
there is an answer as a standard technique 
in the many-body physics.
It is the {\it boson expansion method}.
Using this method, 
we can express the bilocal operators $:M:$
in terms of bosonic operators 
which correspond to $:M:$ in the large $N$ limit.
This method is introduced in the next section 
and applied to the algebra of $:M:$.

%%%%%%%%%%%%%%%%%%%%%%%%%%%%%%%%%%%
%%%%%%          BOSON EXPANSION METHOD           %%%%%%
%%%%%%%%%%%%%%%%%%%%%%%%%%%%%%%%%%%
\section{Application of boson expansion method}

\setcounter{equation}{0}

The boson expansion method \cite{Marshalek} 
is one of the traditional techniques
 in  non-relativistic  many-body problems.
In various systems we encounter a situation 
that there are some boson-like excitations 
even if the fundamental variables of the system 
 are not bosons.
The boson expansion method has been introduced 
for describing such bosonic excitations 
in non-bosonic systems. 
 In this section we briefly review
 this method and apply it to 
our case, QCD$_{1+1}$.

\subsection{Boson expansion method -- a short review}

Originally the boson expansion was invented by 
Holstein and Primakoff \cite{HP}
to describe  interacting spin waves 
in the Heisenberg ferromagnet.
Although the first excitation from the 
spin-aligned vacuum,
 i.e. a spin wave, is solved exactly, 
the next excitation is difficult 
to obtain due to the interaction 
between spin waves, 
which originates from the $SU(2)$ spin algebra.
So they introduced an boson operator 
and represented the spin operators by it 
so that they satisfy the $SU(2)$ algebra.
Then the theory is translated 
into a bosonic theory with interactions.
A single boson state corresponds 
to the single spin wave.
Later  another representation
of the spin algebra was proposed 
by Dyson \cite{Dyson}.
Also, Usui \cite{Usui} applied 
this method to  plasma oscillations.

In nuclear physics, since bosonic excitations from 
the ground state are very important 
for understanding the collective vibrations, 
 the boson expansion method has been 
investigated in great detail 
\cite{Marshalek,BZ,Marumori,Ring-Schuck}. 
It would be  particle-hole operators 
that play an important role for
 describing (collective) bosonic excitations 
in the fermionic many-body problem.
Let $a^{\dagger}_\mu$ be the 
 creation operator of a {\it particle} with 
its quantum number $\mu$
and $d^{\dagger}_i$  that of a  {\it hole}  in state $i$. 
The  four bifermion operators form an algebra
$d_i a_\mu, a_\mu^\dagger d_i^\dagger , 
 d_i^\dagger d_j$ and $a_\mu^\dagger a_\nu$  ;
\begin{eqnarray}
[d_i a_\mu ,a_\nu^\dagger d_j^\dagger] & 
= & \delta_{\mu\nu}\delta_{ij}
-\delta_{\mu\nu}d_j^\dagger d_i
-\delta_{ij}a_\nu^\dagger a_\mu, \label{alg1}\\
{}[d_i a_\mu ,a_\nu^\dagger a_\sigma] &
 = & \delta_{\mu\nu}d_i a_\sigma, \\
{}[a_\mu^\dagger a_\nu, 
a_\sigma^\dagger a_\tau] & = &
 \delta_{\nu\sigma}a_\mu^\dagger a_\tau
-\delta_{\mu\tau}a_\sigma^\dagger a_\nu, \\
{}[d_i a_\mu , d_j^\dagger d_k] &
 = & \delta_{ij}d_k a_\mu , \\
{}[d_i^\dagger d_j,d_k^\dagger d_l] & 
= & \delta_{jk}d_i^\dagger d_l-
\delta_{il}d_k^\dagger d_l .\label{alg5}
\end{eqnarray}
Their first motivation is to obtain collective bosons 
for describing the collective excitation. 
The term "collective excitation" means that the number of the 
participating particles is large.
Thus they usually start with  collective 
fermion-pair operators such as
$
b^{\dagger}_{\mu}=\sum_{\nu i}
C^{\mu}_{\nu i}a_{\nu}^{\dagger}
d_i^{\dagger},
$
and then represent these by bosonic operators \cite{BZ}.
Here however we do not follow the same path 
for  reasons mentioned later
and here we introduce  the bosonic expressions 
for  {\it pure} fermion-pair operators. 
  
There are several ways to implement this algebra.
Here we list three well-known representations.
The different boson expansions  below have been
shown to be  equivalent \cite{Doba}.
Let us begin with the Holstein-Primakoff (HP)
type expansion.
Introducing the boson operators,
\begin{equation}
[B_{\mu i}, B_{\nu j}^\dagger]=\delta_{\mu \nu}\delta_{ij},\ \ 
[B_{\mu i}, B_{\nu j}]=0,\label{Bcom}
\end{equation}
the four operators are represented by 
\begin{eqnarray}
a_\mu^\dagger a_\nu &=& \sum_j B_{\mu j}^\dagger B_{\nu j} 
\equiv {\cal A}_{\nu \mu}, \\
d_k^\dagger d_l &=& \sum_\mu B_{\mu k}^\dagger B_{\mu l},\\
d_i a_\mu &=& \sum_\nu (\sqrt{1-{\cal A}})_{\mu \nu} B_{\nu i},
\label{HPorg3}\\
a_\mu^\dagger d_i^\dagger &=& \sum_\nu B_{\nu i}^\dagger 
(\sqrt{1-{\cal A}})_{\mu \nu}^\dagger.\label{HPorg4}
\end{eqnarray}
These equations should be considered as follows; 
if we substitute the r.h.s. into 
the algebra (\ref{alg1})-(\ref{alg5}), 
they hold by using the bosonic commutators (\ref{Bcom}).
As long as we stay in the zero fermion-number sector,
this replacement  is correct.
The square roots in (\ref{HPorg3}) and (\ref{HPorg4}) 
are defined by the corresponding Taylor series 
(${\cal A}_{\mu \nu}^2=
\sum_{\rho}{\cal A}_{\mu\rho}{\cal A}_{\rho\nu}$). 
The term "expansion" comes from the fact 
that the HP type contains an infinite expansion 
in terms of the boson operator.
There exists a  subspace of  full bosonic Fock space, 
where the square roots are well defined. 
This space is called the {\it physical subspace}.
If we truncate the infinite series by a few terms, 
it would not be a good approximation as it is.
This is why they consider the collective state first 
and apply the boson expansion 
to it instead of the pure bifermion operators. 

Second, there is a finite expansion method, 
the Dyson expansion.
In this method,  the expressions 
 (\ref{HPorg3}) and (\ref{HPorg4}) are
replaced by $d_i a_\mu =  B_{\mu i}$ 
and $ a_\mu^\dagger d_i^\dagger 
= \sum_\nu B_{\nu i}^\dagger 
(1-{\cal A})_{\mu \nu}^\dagger$, respectively.
While the expansion is finite, 
it is evident that the relation 
$(d_i a_\mu )^{\dagger}=a_\mu^\dagger d_i^\dagger$ 
does not hold.
To remedy this, the Dyson expansion needs 
a particular method, 
which is called the unitary projection.

The third representation is the  Schwinger type 
\cite{Schwinger,Blaizot},
which is also finite.
This method needs two kinds of bosons. 
In addition to $B_{\mu i}$ ,
we introduce another boson $A_{ij}$
 which commutes with $B_{\mu k}$;
\begin{eqnarray}
{}[A_{ij},A_{kl}^\dagger]&=&\delta_{ik}\delta_{jl},\ 
\ \ [A_{ij},A_{kl}]=0,  \\
{}[A_{ij},B_{\mu k}]&=&
[A_{ij},B_{\mu k}^{\dagger}]=0.
\end{eqnarray}
Then the four operators are expressed as
\begin{eqnarray}
a_\mu^\dagger a_\nu &=& \sum_j B_{\mu j}^\dagger B_{\nu j} , \\
d_i^\dagger d_j &=& \delta_{ij}-\sum_k A_{jk}^\dagger A_{ik},
\label{Sch2}\\
d_i a_\mu &=& \sum_k A_{ik}^\dagger  B_{\mu k},\\
a_\mu^\dagger d_i^\dagger &=& \sum_k B_{\mu k}^\dagger A_{ik}.
\end{eqnarray}
It is evident from Eq. (\ref{Sch2}) that the state 
annihilated by these bosonic operators is not a true vacuum.
The true vacuum state $|vac>_B$ is given as
\begin{equation}
|vac>_B=\frac{1}{\sqrt{N!}}\sum_{P_h}(-1)^{P_h} P_h 
A^{\dagger}_{1h_1}A^{\dagger}_{2h_2}\cdots A^{\dagger}_{Nh_N}|0),
\end{equation}
where $N$ is the number of the hole states, $P_h$ means 
the permutation of the indices \{$h_1,h_2,\ldots h_N $\},
 and $|0)$ is a state which satisfies $B_{\mu k}|0)=A_{ij}|0)=0$.

\subsection{Boson expansion method -- application to QCD$_{1+1}$}

Let us go back to two-dimensional QCD. 
The algebra  we want to simplify is Eq. (\ref{Algebra}).
Essentially this is the same as that of the
fermion pair operators in  nuclear physics, i.e. 
 Eqs.(\ref{alg1})-(\ref{alg5}). 
The particle-hole operators correspond to 
 quark-antiquark operators, and so on.
The only difference is the existence of  an internal symmetry
($SU(N)$ or $U(N)$) and   continuum indices (momentum). 
Taking these points into account, let us introduce  the bosonic operators,
\begin{eqnarray}
{}[B(p_1,p_2),B^\dagger(q_1,q_2)]&=&\delta(p_1-q_1)\delta(p_2-q_2),\\
{}[B(p_1,p_2),B(q_1,q_2)]&=&0,
\end{eqnarray}
where momentum indices are defined to be positive  $(p_i, q_i>0)$.
Since $:M:$ is constructed so as to be color singlet, 
$B(p_1,p_2)$ need not carry color.
In this paper we consider only HP type expansion.
The bifermion operators are expressed as
\begin{eqnarray}
:M_{-+}(p_1,p_2):&=& \int_0^\infty dq \ \  B^\dagger (-p_1,q) B(p_2,q) 
\equiv {\cal A}(p_2,-p_1), \label{BEM1}\\
:M_{+-}(p_1,p_2):&=& -\int_0^\infty dq\ \  B^\dagger(q,-p_2) B(q,p_1),
\label{BEM2}\\
:M_{++}(p_1,p_2):&=& \int_0^\infty dq\ \  
(\sqrt{N-{\cal A}})(p_2,q)
 B(q,p_1),\label{BEM3}\\
:M_{--}(p_1,p_2): &=& \int_0^\infty dq\ \ B^\dagger (q,-p_2)
(\sqrt{N-{\cal A}})(q,-p_1).\label{BEM4}
\end{eqnarray}
Since $:M_{++}:=O(N^{1/2})$ and $:M_{-+}:=O(N^0)$,
the bosonic operator is considered to be $O(N^0)$.
Of course, the square roots in Eqs. (\ref{BEM3}) and 
(\ref{BEM4}) are defined by 
their Taylor expansions, for example, 
\begin{eqnarray}
:M_{++}(p_1,p_2):&=&\sqrt{N}B(p_2,p_1)\\ \nonumber 
&-& \frac{1}{2\sqrt{N}}\int_0^{\infty}
dq dk B^{\dagger}(q,k)B(p_2,k)B(q,p_1) -\cdots .
\end{eqnarray}
As was noted in the previous section, 
the bosonic representation should satisfy 
two conditions to keep in touch 
with the meson picture.
The first requirement that 
$:M:$ itself becomes bosonic in the large $N$ limit, 
i.e. Eq.(\ref{largeNB})be satisfied as
\begin{equation}
\lim_{N\rightarrow \infty}\frac{1}{\sqrt{N}}:M_{--}(p_1,p_2):
		=B^\dagger(-p_1,-p_2).\label{largeNB2}
\end{equation}
And also the second one that the single two-body state should be 
 bosonic for finite $N$ is observed apparently as
\begin{equation}
:M_{--}(p_1,p_2):|0> = \sqrt{N}B^\dagger(-p_1,-p_2)|0),
\end{equation}
where $|0)$ is a vacuum state of the bosonic operator; 
$B(p,q)|0)=0$.
That is why we used the HP type expansion.

Since  the boson expansion becomes a good approximation
for systems with  large  participating particles,
they start with  collective bosons instead of  
pure particle-hole operators  in nuclear physics \cite{BZ}.
In our case, however, there is a parameter $N$ 
associated with the internal symmetry
and the bilocal operators are color singlet.
The number of  participating particles in the two-body state
goes to infinity if $N\rightarrow \infty$.
Thus we do not have to introduce collective operators at this stage. 
Collective bosons are introduced to give  
local bosons in the next section.
Furthermore, note that also  in the large $N$ limit, 
the effect of the physical subspace vanishes.
The existence of  internal symmetry 
makes the situations simple.

Substituting (\ref{BEM1})-(\ref{BEM4}) 
into the Hamiltonian (\ref{Ham}),
we obtain for the leading contribution; 
\begin{eqnarray}
H_{HP}^{(0)}&=&-\frac{g^2N}{8\pi}
\int_{0}^{\infty}dp_1dp_2dq_1dq_2\ 
\delta(p_1+p_2-q_1-q_2)\times \nonumber  \\
&&\hspace{1cm}\frac{{\cal P}}{(p_1-q_2)^2}
B^{\dagger}(p_1,p_2)B(q_2,q_1)\nonumber  \\
&&+\frac{1}{2}\left(m^2-\frac{g^2N}{2\pi}\right)
\int_{0}^{\infty}dq dq'\ 
\left( \frac{{\cal P}}{q}+\frac{{\cal P}}{q'}\right)
B^{\dagger}(q,q')B(q,q'),
\end{eqnarray}
where the suffices (0) and HP imply the  $O(N^0)$ 
contribution and the Holstein-Primakoff type expansion.
The equation of motion is 
\begin{eqnarray}
i\partial_+B(p,r-p)&=&[B(p,r-p), H_{HP}^{(0)}] \nonumber  \\
&=&-\frac{g^2N}{4\pi}\int_0^r dq \frac{{\cal P}}{(p-q)^2}
B(q,r-q) \nonumber  \\
&& +\frac{1}{2}\left(m^2-\frac{g^2N}{2\pi}\right) 
\left( \frac{{\cal P}}{p}+\frac{{\cal P}}{r-p}\right)B(p,r-p).
\end{eqnarray}
Essentially this is equivalent to 
the 't Hooft equation (\ref{tHoofteq}).

The higher order Hamiltonians ($H_{HP}^{(n)}=O(N^{-n/2})$) 
are obtained by substituting the expansion of 
Eqs. (\ref{BEM1})-(\ref{BEM4}). 
The next leading Hamiltonian is on the order of $N^{-1/2}$
 and is given, after normal ordering, by
\begin{eqnarray}  
H_{HP}^{(1)}&=&\frac{g^2\sqrt{N}}{4\pi}
\int_0^{\infty} dp_1dp_2dq_1dq_2dk\ \
\delta(p_1-p_2+q_1+q_2)\ \
 \frac{{\cal P}}{(p_1+q_2)^2}\times \nonumber \\
& &\left\{ B^{\dagger}(k,p_2)B(k,p_1)B(q_2,q_1)
-B^{\dagger}(p_2,k)B(p_1,k)B(q_1,q_2) + {\rm h.c.}   \right\},
\end{eqnarray}
which corresponds to a three-point vertex.

The expressions (\ref{BEM1})-(\ref{BEM4}) are very similar 
to those  of  Nakamura and Odaka \cite{Naka-Oda}.
They introduced bosonic operators 
as the leading contribution of $:M:$ in the $1/N$ expansion. 
This construction is the same as the requirement (\ref{largeNB2})
and indeed their bosonic operator is equivalent  
to ours in the large $N$ limit.
However, if we express their representation in our notation, 
it is different from ours;
\begin{eqnarray}
:M_{-+}(p_1,p_2):&=& \int_0^\infty dq \ \ 
 B^\dagger (-p_1,q) B(p_2,q) ,\nonumber  \\
:M_{+-}(p_1,p_2):&=& -\int_0^\infty dq\ \  
B^\dagger(q,-p_2) B(q,p_1)
\equiv {\cal C}(p_2,-p_1), \nonumber \\
:M_{++}(p_1,p_2):&=& \int_0^\infty dq\ \  
(\sqrt{N+{\cal C}})(p_2,q)
 B(q,p_1),\nonumber \\
:M_{--}(p_1,p_2): &=& \int_0^\infty dq\ \ B^\dagger (q,-p_2)
(\sqrt{N+{\cal C}})(q,-p_1). \nonumber
\end{eqnarray}
This is one of the HP type expansions.
They were not aware that 
this is the boson expansion methods. 
However, once we recognize the usefulness 
of the boson expansion method, 
we can apply this idea to many other models even with small $N$.
Other $finite$ expansion methods such as the Dyson type 
or methods using collective fermion pairs
will  be found to be more useful for such theories.

%%%%%%%%%%%%%%%%%%%%%%%%%%%%%%%
%%%%%%%%       LOCAL BOSONS       %%%%%%%%%%
%%%%%%%%%%%%%%%%%%%%%%%%%%%%%%%

\section{Local boson operators as collective states}
\setcounter{equation}{0}

It was bilocal operators that we introduced 
in the previous section. 
Here  we construct local boson operators.
The following is essentially the same as that of 
Nakamura and Odaka \cite{Naka-Oda}.
However, according to the many-body physics, 
their construction is understood as the introduction 
of the collective states from $B(p,q)$.
Collective boson operators are defined as the linear combination 
of  bilocal bosons with their total momentum  $r$,
\begin{equation}
b_r^{\dagger}= \sqrt{r}\int_0^1 dz\ \ 
 \phi(z) B^{\dagger}(rz,(1-z)r).
\end{equation}
Let us assume that the wave function $\phi(z)$ 
forms an orthonormal complete set;
\begin{eqnarray}
\int_0^1 dz\ \  \phi_n^*(z)\phi_m(z)=\delta_{nm},\\
\sum_n \phi_n(z)\phi_n^*(z')=\delta(z-z').
\end{eqnarray}
As far as $\{ \phi_n\}$ is a complete set, the operators 
\begin{eqnarray}
b_r^{(n)\dagger}&=& \sqrt{r}\int_0^1 dz\ \ 
 \phi_n(z) B^{\dagger}(rz,(1-z)r),\\
b_r^{(n)}&=&\sqrt{r} \int_0^1 dz\ \ 
 \phi_n^*(z) B(rz,(1-z)r),
\end{eqnarray}
satisfy the bosonic commutators
\begin{equation}
[b_r^{(n)}, b_{r'}^{(m)\dagger}]=\delta_{nm}\delta (r-r'), \ \ \ 
[b_r^{(n)},b_{r'}^{(m)}]=0.
\end{equation}
Let us determine $\{ \phi_n\}$ so that $b$ and $b^{\dagger}$ 
diagonalize the lowest Hamiltonian $H_{HP}^{(0)}$.
Inserting these operators, the equation 
of motion can be written as 
\begin{equation}
\mu^2_n \phi_n(z)=\left(m^2-\frac{g^2N}{2\pi}\right)
\left(\frac{1}{z}+\frac{1}{1-z}\right)\phi_n(z)
-\frac{g^2N}{2\pi}\int_0^1 dz' \frac{\phi_n(z')}{(z-z')^2},
\end{equation}
which means  that the wave function $\phi_n(x)$ 
is a solution of the 't Hooft equation.
If we define a scalar field by 
\begin{equation}
\Phi_n(x)\equiv \frac{1}{\sqrt{2\pi}}
\int_0^{\infty}\frac{dr_-}{\sqrt{2r_-}}
\ \ \left( b_r^{(n)} e^{-irx}+b_r^{(n)\dagger } e^{irx}  \right),
\end{equation}
this is a field operator of the $n$-th excitation state.
$\Phi_n(x)$ satisfies the usual LF commutator;
\begin{equation}
[\Phi_n(x),\ \Phi_m^{\dagger}(y)]_{x^+=y^+}=
-\frac{i}{4}\delta_{nm}\epsilon (x^--y^-).
\end{equation}
Using this operator, $H_{HP}^{(0)}$ 
is diagonalized and can be written as 
\begin{eqnarray}
H_{HP}^{(0)}&=&\sum_n \int_0^{\infty}dr\ \ 
\frac{\mu_n^2}{2r}\ \ b_r^{(n)\dagger} b_r^{(n)}\nonumber \\
&=&\sum_n\int_{-\infty}^{\infty}dx\ \ \frac{\mu_n^2}{2}:\Phi_n(x)^2:.
\end{eqnarray}
This is the Hamiltonian of  free scalar bosons with their mass being the 
eigenvalues of the 't Hooft equation.
The next order Hamiltonian is 
\begin{equation}
H_{HP}^{(1)}=\frac{g^2\sqrt{N}}{4\pi}
\int_0^{\infty}\frac{dR}{\sqrt{R}}
\int_0^1 dz \sum_{nml}\left( K_{nml}(z)-\tilde{K}_{nml}(z)   \right)
\left( b_R^{(n)\dagger}b_{Rz}^{(m)}b_{R(1-z)}^{(l)} + {\rm h.c.} \right),
\end{equation}
where 
\begin{eqnarray}
K_{nml}(z)&=&\int_0^1 dx \int_0^1 dy
 \frac{\sqrt{z(1-z)}}{\{z(1-x)+(1-z)y\}^2}
\ \ \phi_n^*(zx)\phi_m(x)\phi_l(y),\nonumber \\
\tilde{K}_{nml}(z)&=&\int_0^1 dx \int_0^1 dy 
\frac{\sqrt{z(1-z)}}{\{zx+(1-z)(1-z)\}^2}
\ \ \phi_n^*(1-z(1-x))\phi_m(x)\phi_l(y). \nonumber 
\end{eqnarray}
Hence for finite $N$, we obtain 
a Hamiltonian of infinite kinds of bosons interacting with each other.
In principle, there are infinitely many-point vertices which 
originate from the infinite expansion of the HP type.

Recently Barb\'{o}n and Demeterfi \cite{Barbon} 
derived an effective Hamiltonian  
in the approximation that the following operators are bosonic; 
\begin{eqnarray}
\alpha_P^{\dagger}&\equiv&{\cal N}
\int_0^P dk_-\phi(k_-):M_{--}(-k_-,-P+k_-):\ \ ,\\
\alpha_P&\equiv&{\cal N}\int_0^P 
dk_-\phi^*(k_-):M_{++}(P-k_-,k_-):\ \ .
\end{eqnarray}
Indeed they satisfy $[\alpha_P, \alpha_Q^{\dagger}]=
\delta(P-Q)+ O(1/N)$
and thus are  bosonic in the large $N$ limit, 
which has been insisted  on many times.
It is characteristic that their Hamiltonian 
has only finite terms in contrast to ours.
This is clearly because they did not incorporate 
the next leading order of the commutator.  
Their method cannot be extended to small $N$ theories, 
while the boson expansion methods have other ways 
to avoid it, as has been noted.

%%%%%%%%%%%%%%%%%%%%%%%%%%%%%%%%%
%%%%%%%%%%      CONCLUSION     %%%%%%%%%%%%
%%%%%%%%%%%%%%%%%%%%%%%%%%%%%%%%%
\section{Conclusion and Discussions}
\setcounter{equation}{0}

In this paper we have found that  the boson expansion methods
are useful for constructing bosonic theories from two dimensional QCD 
on the light-front. 
In this method, fermion bilinear operators can be nonlinearly 
expressed by bilocal bosonic operators such that the same algebra holds.
Among various representations, 
if we choose the Holstein-Primakoff type  expansion,  
the bosons are identified with mesons in the large $N$ limit, 
and for finite $N$ we obtain an interacting boson theory.
The physical meaning of this procedure is as follows.
Because of the inevitable effects of the fermion statistics,
the two-body states cannot stay bosonic.
This effect is called the kinematical effect because
it is independent of the Hamiltonian.
The kinematical complexity in the two-body states is 
translated into interactions among bosons 
via the boson expansion methods. 
Furthermore, we can build local boson operators from 
bilocal bosons as the collective state.
Since this is determined to make the  lowest Hamiltonian diagonal,
 this effect is called the dynamical effect.
Eventually we have obtained a  local interacting 
theory  with  infinite  kinds  of bosons.
Thanks to the existence of the parameter $N$,
we can argue the above two effects separately, 
which is not the case in nuclear physics.

In fact, the introduction of the gauge invariant operator 
was not crucial for implementing the boson expansion method, 
but it has made the argument very transparent and 
leads to the discovery of the applicability 
of the boson expansion methods. 
Although there are no dynamical degrees of freedom 
in the gauge field when  the  space is infinite,
if we work in finite (light-front) space, 
there exist dynamical zero modes of the gauge field.
The gauge-invariant operators contain them
which is not the case in nuclear physics.
The boson expansion method can also be applied to these 
operators as far as the winding of the zero mode is trivial.

The application of the boson expansion method 
has been performed for  light-front field theory.
However, we can do the same things for 
 perturbation theory in an 
equal-time formulation.
The merit on the light-front is, 
thanks to the triviarity of the vacuum,  that
the few-body states are  expected to be good variables 
for describing hadrons.
This is true even for small $N$ theories.
Thus we expect that  bosons also describe hadrons.
In  equal-time formulations, however, 
we cannot expect in general that  
only the two-body state gives a meson.
Thus  boson expansion methods are expected to 
work better on the light-front than on the equal-time.

There are many things to be clarified in this formalism.
Especially, it is important to know the relation to
 the usual bosonization.
In  bosonization, the number of  bosons is finite,
while it is infinite in our formalism. 
As a more concrete problem, the existence of  
meson-meson bound states is suggested  in Ref. \cite{Sugihara}.
It would be interesting to investigate it in our framework.
Furthermore, there is the problem of how we can describe baryons.
Recently Rajeev \cite{Rajeev} argued that the bilocal operator 
$M(p,q)$ is an element of an infinite dimensional Grassmannian,
and that the baryon number is understood as the topological invariant 
(virtual rank) of the Grassmannian, and thus
 baryons are interpreted as solitons. 
One of our motivations was to see
 how we can obtain an effective theory such as the Skyrme model.
In the Skyrme model, a baryonic state is considered as a soliton
in the large $N$ limit \cite{Witten}. 
Our resulting Hamiltonian  is a {\it nonlinear} theory of mesons. 
Although the topological effect  is difficult  to 
observe in the trivial vacuum of the light-front,
it is intriguing to find a solitonic configuration in our formalism.

In nuclear physics,
the boson expansion method is considered as a method that
goes beyond the Tamm-Dancoff approximation. 
Since the structure of the LF theories is similar
 to  non-relativistic quantum theory,
it seems to be natural that we have found that
the boson expansion method 
is also applicable to LF field theory 
when we want to go beyond the TD approximation.
Also, in  many-body physics, 
the boson introduced is thought to be appropriate for
describing the Nambu-Goldstone
boson when there is symmetry breaking.
Indeed at the very first application, 
the spin wave corresponds to 
the NG boson under  spontaneous magnetization.
Therefore there might be a possibility that also on the LF 
the boson  introduced in the boson expansion methods 
can describe the NG boson.
Since this is a very important problem, 
we should apply the boson expansion method 
to higher dimensional theories where the NG boson or 
the pseudo NG boson will exist.

%%%%%%%%%%%%%%%%%%%%%%%%%%%%%%%%%
%%%%%%%%      ACKNOWLEDGEMENT        %%%%%%%%
%%%%%%%%%%%%%%%%%%%%%%%%%%%%%%%%%

{\bf Acknowledgments}

The author would like to thank Prof. K. Ohta for discussions
and Dr. H. Suganuma and Dr. Ichinose
for continuous encouragement.
The hospitality of the particle theory group in the Kyushu University,
where part of this work was done,
is also acknowledged.

%%%%%%%%%%%%%%%%%%%%%%%%%%%%%%%%%
%%%%%%%%%           APPENDIX             %%%%%%%%%%%
%%%%%%%%%%%%%%%%%%%%%%%%%%%%%%%%%
\appendix
\section{}
If we use the fermion operator after solving the constraints, 
we can easily relate the bilocal operators $:M:$ to the quark 
or anti-quark operators.
As in eq.(\ref{usualrep}), 
the usual definition of the quark or anti-quark 
creation/annihilation operators is given by
\begin{equation}
\psi_i(x)=\int_0^\infty \frac{dk_-}{\sqrt{2\pi}}
\left\{ a^{i}_{k_-} e^{-ikx}+d_{k_-}^{i\dagger} e^{ikx} \right\},
\nonumber
\end{equation}
where the momentum takes only positive values $k_->0$.
On the other hand, Fourier transformation of 
the gauge invariant operators 
is given as in eq. (\ref{bilocal_mom}).
Thus its momenta can take negative values.
Taking care of this point, the relation between two is
given as follows:
\begin{eqnarray}
:M_{++}(p,q):&=& d_p^ia_q^i   \\
:M_{--}(p,q):&=& a_{-p}^{i\dagger}d_{-q}^{i\dagger}   \\
:M_{+-}(p,q):&=&-d_{-q}^{i\dagger}d_p^i    \\
:M_{-+}(p,q):&=&  a_{-p}^{i\dagger}a_q^i  
\end{eqnarray} From 
these we can see that $:M_{++}:$ is the 
 hermitian conjugate to $:M_{--}:$,
\begin{equation}
\left(   :M_{++}(p,q):    \right)^{\dagger}=:M_{--}(-q,-p):\ \ .
\end{equation}

%\newpage

%%%%%%%%%%%%%%%%%%%%%%%%%%%%%%%%
%%%%%%%%%         REFERENCES      %%%%%%%%%%%
%%%%%%%%%%%%%%%%%%%%%%%%%%%%%%%%

\end{document}